\documentclass{proc}
\usepackage{amsfonts}
\usepackage{amsmath}

\usepackage{latexsym}
\usepackage{amsmath}
\usepackage{amssymb}
\usepackage{subfigure}
\usepackage{algorithm}

\usepackage{algorithm,algpseudocode}
\usepackage{amssymb}
\usepackage{amsmath}

\usepackage{cite}
\usepackage[export]{adjustbox}

\usepackage{amssymb}
\usepackage{amsmath}

\usepackage{cite}
\usepackage{algorithm}
\usepackage{algpseudocode}
\usepackage{tablefootnote}

\usepackage{amssymb}
\usepackage{amsmath}
\usepackage{epstopdf}
\usepackage{cite}
\usepackage{algorithm}
\usepackage{algpseudocode}
\usepackage{tablefootnote}
\newtheorem{theorem}{Theorem}[section]

\newcommand{\BEQA}{\begin{eqnarray}}
\newcommand{\EEQA}{\end{eqnarray}}

\newcommand{\define}{\stackrel{\triangle}{=}}
\newtheorem{remark}[theorem]{Remark}

\begin{document}

\title{Identification of Source of Rumors in Social Networks with Incomplete Information}
\author{A.~Louni$^{1}$, S.~Anand$^{2}$, K.~P.~Subbalakshmi$^{1}$ \\
$^{1}$Department of ECE, Stevens Institute of Technology
Hoboken, NJ 07030\\
$^{2}$Department of Electrical Engineering, New York Institute of Technology,
New York, NY 10019\\
alouni@stevens.edu, asanthan@nyit.edu, ksubbala@stevens.edu}
\maketitle

\begin{abstract}
Rumor source identification in large social networks has received significant attention lately. Most recent works deal with the scale of the problem by observing a subset of the nodes in the network, called sensors, to estimate the source. This paper addresses the problem of locating the
source of a rumor in large social networks where some of these sensor nodes have failed. We estimate the missing information about the sensors using doubly non-negative (DN) matrix completion and compressed sensing
techniques. This is then used to identify the actual source by using a maximum likelihood estimator we developed earlier, on a large data set from
Sina Weibo. Results indicate that the estimation techniques result
in almost as good a performance of the ML estimator as for
the network for which complete information is available. To the
best of our knowledge, this is the first research work on source
identification with incomplete information in social networks.

\end{abstract}

\section{Introduction}
On April $2013$, hackers took control of the Twitter account @AP and sent a fake tweet  about explosions in the White House.
U.S. financial markets were spooked by this tweet;
  the  index value of S\&P $500$    dropped $14$ points, wiping out $\$136.5$ billion  in a matter of seconds  before the financial markets recovered\cite{stock13}. At a time when cybersecurity  has become a major national issue,   the ease of rumor spread through social networks has exacerbated concerns.
	More specifically, studies show that rumors spread much faster in social networks than other type of networks, even faster than networks with complete graph topology\cite{Doer11}. Therefore,  it is of great interest to pinpoint the source of the rumor in  time  by leveraging the social network topology and observing  the state of nodes. The practical applications include rapid damage control  and understanding the role of network structure in rumor dissemination, thereby facilitating the design of sophisticated policies to prevent further viral spreading of misinformation through social networks in the future.
	
The various approaches in locating the source of a rumor  may
be classified based on whether they rely
on observing all the nodes in the social network \cite{Zaman1,Zaman3,Luo13,Prakash12, Zhu13, Luo2013, Dong13} or a fraction of nodes in the social network\cite{Pinto12,infocom}. It is impractical to observe all the nodes in the social network due to the large amount of the computational complexity that is involved.  One means to deal with the complexity issues is  by selecting a subset of nodes (also called sensors) \cite{Pinto12,infocom}.
In \cite{Pinto12},  a maximum likelihood (ML) estimator  was proposed using measurements by the sensors.
It was shown  that an average
source localization error of less than $4$ hops can
be achieved by observing $20\%$ of the nodes in network.
In \cite{infocom} we proposed a two-stage source localization
algorithm that required $3\%$ less sensor nodes to provide measurements on the time of arrival of information, and yet provided results with
the same accuracy as previous studies.

In most practical scenarios, it is not possible to observe the status of all nodes in a large-scale social network.
Therefore, the source of rumor must be located based on the
measurements  collected  by a subset of
nodes (called sensors) in the social network.
 The sensors record the arrival
times of the rumor to estimate the most likely source.  However, in most practical scenarios, we may not have  complete information on the time at which the sensors receive the rumor. This could happen because  most social networks such as Twitter do
not provide  public access to their full stream
of tweets and many Facebook users keep their activity and profiles, private.    Overall, the rapid growth of the social networks themselves, and the increasing volume of
their generated data, will likely  augment the problem of missing data in the study of rumor diffusion.     This
paper presents a technique to locate the source
of a rumor for large social networks where the  information on the time at which the sensors receive the rumor is incomplete.

Using incomplete information to estimate the source of rumors is achieved by recovering the missing information. Such  data recovery was addressed  in the context of computer networks \cite{elseviercomnets} and   sensor networks \cite{Fazel12}. In the context of social networks, recovering the missing information is essentially a matrix completion problem. There are several  approaches to matrix completion \cite{dono06,cand09,matrixcompletion,Hogb02,drew98}. We deploy compressed sensing \cite{dono06,cand09} and doubly non-negative (DN) matrix completion \cite{matrixcompletion,Hogb02,drew98} to recover the missing information on the time epochs at which certain sensors receive the information. We also present a renewal theory-based argument  to improve the DN completion based estimation.
We use the estimated values to identify the source of rumors using a maximum likelihood (ML) estimator we developed in \cite{infocom}.
Results indicate that these estimation methods provides us with almost as good a performance of  source identification as that when complete information is available.

The rest of the paper is organized as follows.
In Section~\ref{sec:sourceid}, the
rumor diffusion model and the source estimator are
discussed.  Section~\ref{sec:partial}  presents different approaches to recover missing
values at the sensors using compressed sensing, DN matrix completion, and renewal theory-based model.  Experimental
results are provided in Section~
\ref{sec:results} and conclusions in Section~\ref{sec:conc}.

\section{Source Identification}\label{sec:sourceid}

The source identification mechanism we designed in \cite{infocom} is as follows\footnote{The details can be found in \cite{infocom} but we present the key results here to enable easier reading of this paper for the reader.}. A social network can be modeled as a graph, $G(V,E)$, where a vertex, $v\in V$ represents a user in the network and two
vertices, $u$, $v\in V$ share an edge if the corresponding users share a friendship or any similar relation.
Whenever a user tweets or posts a message (or a rumor), the people following the user or the friends of the user may re-tweet or re-post the rumor. Let $t_{mn}$ be the delay between the epochs at which nodes,
$m$ and $n$ get ``infected''\footnote{By ``infected'', we mean that a node not only receives a rumor but also re-posts or re-tweets because he/she believes the rumor.} by a rumor. Then $t_{mn}\sim
\mathcal{N}(\mu_{mn},\sigma_{mn}^2)$ \cite{Pinto12}. The parameters, $\mu_{mn}$ and $\sigma_{mn}^2$ depend on the path between the nodes, $m$ and $n$.
A source, $s^*$,  starts a rumor and spreads it on a social network. The information diffuses through the network and reaches nodes, $v\in V$ along the shortest path from $s$ to $v$. The goal is to determine $s^*$ given the time epochs at which nodes, $v \in S\subset V$ (the set of sensor nodes) receive the information.

The source localization algorithm consists of two stages. In the first stage, the
cluster that  most likely contains the source
of the rumor is identified and then, in the second stage, we search within this cluster and identify the source of the rumor. A new graph
$G^{\rm gate}=(V^{\rm gate},E^{\rm gate})$, where $V^{\rm gate}$ is the gateway nodes (nodes connecting clusters using between-cluster ties), $E^{\rm gate}$ is  incident on the vertices in $V^{\rm gate}$.
Let $S=\{l_1,l_2,...,l_{k_{1}}\}$ be a set of $k_1$ nodes, selected from $V_{\rm gate}$, to observe the time arrival of the rumor.
  Since the time that the source starts to spread information, ${t^*}$, is typically unknown, inter-arrival times,  ${\Delta {t_i}} \buildrel \Delta \over = ({t_i}+{t^*}) - ({t_1}+{t^*})={t_i}-{t_1}$, can be used for estimation, where $t_i$ is the time at which the rumor is received at the $i^{\mathrm{th}}$ sensor in $G^{\rm gate}$.
	The  inter-arrival time observation vector is then defined as ${\mathbf{\Delta {t^{stage1}}}} =[\Delta {t_2},\Delta {t_3},...,\Delta {t_{k-1}}]^T$ \footnote{$(.)^T$ represents the transpose of a vector or a matrix.}.
	Since, all the nodes are equally likely to be the source of a rumor,
 the maximum likelihood (ML) estimator is the optimal estimator for the source of the rumor, described as

\begin{equation}
\label{eq43}
\begin{split}
\Hat v^{(1)} &=\\
&\mathop {\arg {\rm{ max}} }\limits_{v \in {V^{gate}}} ~\frac{1}{{{(2\pi )^{\frac{{{k_1}- 1}}{2}}}\det {{({\mbox{\boldmath$\Lambda$} _{v}})}^{1/2}}}}\times \\
&\exp ( - \frac{1}{2}{(\mathbf{\Delta {t^{stage1}}} - {\mbox{\boldmath$\mu$}_{v}})}{({\mbox{\boldmath$\Lambda$} _{v}})^{ - 1}}(\mathbf{\Delta {t^{stage1}}} - {\mbox{\boldmath$\mu$}_{v}})^T),
\end{split}
\end{equation}
where $ {\mbox{\boldmath$\mu$}_{v}}(r)$ is the mean value of difference in arrival times between the first
and the $(r+1)^{\mathrm{th}}$ sensors and
 $\mbox{\boldmath$\Lambda$} _{v} (a,b)$ is the cross-correlation matrix of  difference in arrival times between the $a^{\mathrm{th}}$ and the $b^{\mathrm{th}}$ sensors.

%

In the second stage, the search space will be limited to the nodes inside the cluster that is associated with $\Hat v^{(1)}$.  Let   $G^{\rm cluster} = (V^{\rm cluster},E^{\rm cluster},\mathbf{w}^{\rm cluster})$ be the graph of the nodes inside the most likely candidate cluster. $k_2$ sensors are employed at this stage to collect information about the rumor.
 The corresponding
 optimal ML estimator  is given by
\begin{equation}
\label{eq578}
\begin{split}
{\Hat v^{(2)}} =&\\
 &\mathop {\arg {\rm{ max}} }\limits_{v \in {V^{cluster}}} ~ \frac{1}{{\det {{({\mbox{\boldmath$\Lambda$} _{v}})}^{1/2}}}} \\
&\exp ( - \frac{1}{2}{(\mathbf{\Delta {t^{stage2}}} - {\mbox{\boldmath$\mu$}_{v}})}{({\mbox{\boldmath$\Lambda$} _{v}})^{ - 1}}(\mathbf{\Delta {t^{stage2}}} - {\mbox{\boldmath$\mu$}_{v}})^T)
\end{split}
\end{equation}
where  $\mathbf{\Delta {t^{stage2}}}$ is the observation vector in the second stage and ${\Hat v^{(2)}}$ is the estimated source of the rumor. Detailed information about the two-stage algorithm can be found in \cite{infocom}. It is observed that the ML source estimator requires full information about the vectors, $\mbox{\boldmath{$\Delta$}}\mathbf{t}^{\rm stage 1}$ and $\mbox{\boldmath{$\Delta$}}\mathbf{t}^{\rm stage 2.}$
In most practical scenarios, we may not have the entire
information on the time epochs at which different sensors
receive the information. This could be because most social
networks such as Twitter do not provide public access to
their full stream of tweets and most Facebook users keep
their activity and profiles private. In the following section, we present estimation mechanisms that enable source identification when incomplete information about $\mbox{\boldmath{$\Delta$}}\mathbf{t}^{\rm stage 1}$ and $\mbox{\boldmath{$\Delta$}}\mathbf{t}^{\rm stage 2.}$ is available.

\section{Source Identification with Partial Information}\label{sec:partial}
We present three
different approaches to recover the missing information,
which, in turn, will be used in the analysis to identify
the source, as detailed in Section \ref{sec:sourceid}. First,
we present a compressed sensing based approach (Section \ref{subsec:compressedsensing})
that is effective in recovering sporadically missing information.
Then we present an approach using doubly non-negative (DN)
matrix completion to recover information missing in bursts,
in Section \ref{subsec:DNcompletion}. Then, in Section \ref{subsec:renewal},
we improve the DN completion mechanism by using a renewal
theory-based analysis.
\subsection{Compressed Sensing}\label{subsec:compressedsensing}

Consider a observation vector~$\mathbf{\Delta t}=[\Delta{t_1},\Delta{t_2},...,\Delta{t_K}]^T$,  where $\Delta{t_i}$
corresponds to difference in arrival times between the $i^{th}$ sensor and the reference sensor.
Let $y \in \mathbb{R}^{L}$ be the vector of available entries in $\mathbf{\Delta t}$ where $L \le K$. Hence,
\begin{equation}
\label{eq:1}
\mathbf{y} = \phi \mathbf{\Delta t}
\end{equation}
where $\phi$
is an $L \times K$ measurement matrix.  In order to recover the original observation vector $\mathbf{\Delta t}$ from $\mathbf{y}$, we assume  that there exists an
invertible $K \times K$ sparsifying matrix $\psi$ such that
\begin{equation}
\label{eq:2}
\mathbf{\Delta t} = \psi \mathbf{x}
\end{equation}
where $\mathbf{x} \in \mathbb{R}^{K}$
is   M-sparse with $M \le L$, i.e., it has
only $M$ non-zero entries.   Using Eqn.(\ref{eq:1}) and Eqn.(\ref{eq:2}) we can write
\begin{equation}
\label{eq:3}
\mathbf{y} = \phi \mathbf{\Delta t} = \phi \psi \mathbf{x} = \theta \mathbf{x}
\end{equation}
that is, in general, an ill-posed and ill-conditioned with
$\theta  = \phi \psi$ of dimensions $L \times K$.
Infinitely many solutions are
possible unless we impose some additional constraints on
$\mathbf{\Delta t}$. Since the rumor spread along the shortest paths in the social network, the observation vector shows some amount of correlation among its elements. The correlation structure of the observation vector makes it possible to acquire sufficiently accurate representations of the observation vector without collecting time arrivals from each sensor node. Therefore, the vector $\mathbf{\Delta t}$  can be
approximated by  a low-rank vector $\mathbf{x}$. Therefore, the problem becomes

\begin{equation}
\label{eq:comp2}
\begin{aligned}
& \underset{\mathbf{x}}{\text{min}}
& &{\left\| \mathbf{x} \right\|_{{\ell_1}}} \\
& \text{s.t.}
& & \mathbf{y} = \theta \mathbf{x}
\end{aligned}
\end{equation}
where ${\left\| \mathbf{x} \right\|_{{\ell_1}}}$ is the $\ell_1$-norm of $\mathbf{x}$.

 Ultimately the estimated time arrival vector $t_\textrm {est}$ is
\begin{equation}
\mathbf{\Delta t}_\textrm{est} = \psi \mathbf{x}_{\textrm{opt}}
\end{equation}
where $\mathbf{x}_{\textrm{opt}}$  is the solution to the problem in Eqn.~\ref{eq:comp2}.

\subsection{DN completion}\label{subsec:DNcompletion}
There are certain scenarios, where in the information
on the time epochs of information arrival at different
users corresponding to sensor nodes, may be missing in
bursts. This could happen because certain users that act
as sensors, remain idle, temporarily.  Let $X_{ij}$ denote the time delay between
the epochs when information propagated by sensor node $i$
and that when it reaches node $j$. The delay between different
nodes can then be written as a matrix, $\mathbf{D}=
\left [X_{ij}\right ]_{i,j\in \mathbf{S}}$, where $\mathbf{S}$
is the set of all sensor nodes. Note that in $\mathbf{D}$,
$X_{ii}=0$, $\forall$ $i$. \\
When some nodes temporarily get de-activated or unsubscribe
as a sensor, then the matrix, $\mathbf{D}$, has certain entries
missing. If the set of missing entries occur in bursts, then
techniques like matrix completion can be used to determine the
missing entries in the matrix, $\mathbf{D}$ \cite{matrixcompletion}.
These include inverse $\mathcal{M}-$matrix completion \cite{Hogb02},
doubly non-begative (DN) completion \cite{drew98}.
In order to perform these matrix completions, it is essential
that the matrix, $\mathbf{D}$ represented as a graph\footnote{In
this graph, each row or column represents a vertex and two vertices
are joined by an edge if the value of the corresponding entry is
known.} the graph forms a block clique \cite{DNproperty}. Then
$\mathbf{D}$ is symmetric and of the form
\BEQA\label{eqn:dform}
\mathbf{D}=
\left (
\begin{array}{ccc}
\mathbf{A} & \mathbf{c} & \mathbf{X}\\
\mathbf{c}^T & e & \mathbf{d}^T\\
\mathbf{X}^T & \mathbf{d} & \mathbf{B}
\end{array}
\right ),
\EEQA
where $e$ is any constant, $\mathbf{A}$ is a known $m\times m$ sub-matrix,
with entries $a_{ij}=X_{ij}$, the time delay between
pairs of the first $m$ sensors, $\mathbf{c}$ is an
$m\times 1$ vector, $\mathbf{d}^T$ is a $1\times n$
vector and $\mathbf{B}$ is another $n\times n$ sub-matrix.
An optimal mechanism for DN completion of such a matrix
was discussed in \cite{elseviercomnets}, which we use here
to estimate the missing delays between the times at which
different sensor nodes obtain information. According to
the analysis in \cite{elseviercomnets},
\BEQA\label{eqn:DNcomp}
\mathbf{X}=\mathbf{D}_{\alpha}\mathbf{c}\mathbf{d}^T\mathbf{D}_{\beta},
\EEQA
where $\mathbf{D}_{\alpha}=$diag$(\alpha_1,\alpha_2,\cdots,\alpha_m)$
and $\mathbf{D}_{\beta}=$diag$(\beta_1,\beta_2,\cdots,\beta_n)$. Based
on the values of $E(\mathbf{X})=E\left (\left [X_{ij}\right ]_{1\leq i\leq m
\atop 1\leq j\leq n}\right )$, the optimal values of $\mbox{\boldmath{$\alpha$}}=
\left [ \alpha_i\right ]_{1\leq i\leq m}$ and $\mbox{\boldmath{$\beta$}}=
\left [ \beta_j\right ]_{1\leq j\leq n}$, that minimize the mean square
error in estimation (i.e., the MMSE estimate \cite{srinath}), is given
by the iterative set of equations \cite{elseviercomnets}
\BEQA\label{eqn:estalpha}
\mbox{\boldmath{$\alpha$}}=\frac{E(\mathbf{X}) \mbox{\boldmath{$\beta$}}}
{| |\mbox{\boldmath{$\beta$}}| |^2}
\EEQA
\BEQA\label{eqn:estbeta}
\mbox{\boldmath{$\beta$}}=\frac{E(\mathbf{X}^T) \mbox{\boldmath{$\alpha$}}}
{| |\mbox{\boldmath{$\alpha$}}| |^2}.
\EEQA
It was shown in \cite{elseviercomnets} that the set of
iterative equations in Eqns. (\ref{eqn:estalpha}) and (\ref{eqn:estbeta})
converge if the condition number of the co-variance matrix
of $\mathbf{X}$ is less than $2$. This can be satisfied by adding
sufficiently large values to the diagonal elements of $\mathbf{D}$
(i.e., have $d_{ii}$ as a very large value instead of $0$).
\subsection{Renewal Theory-based Model}\label{subsec:renewal}
\begin{figure}[htbp]
	\centering
		\includegraphics[width=0.5\textwidth]{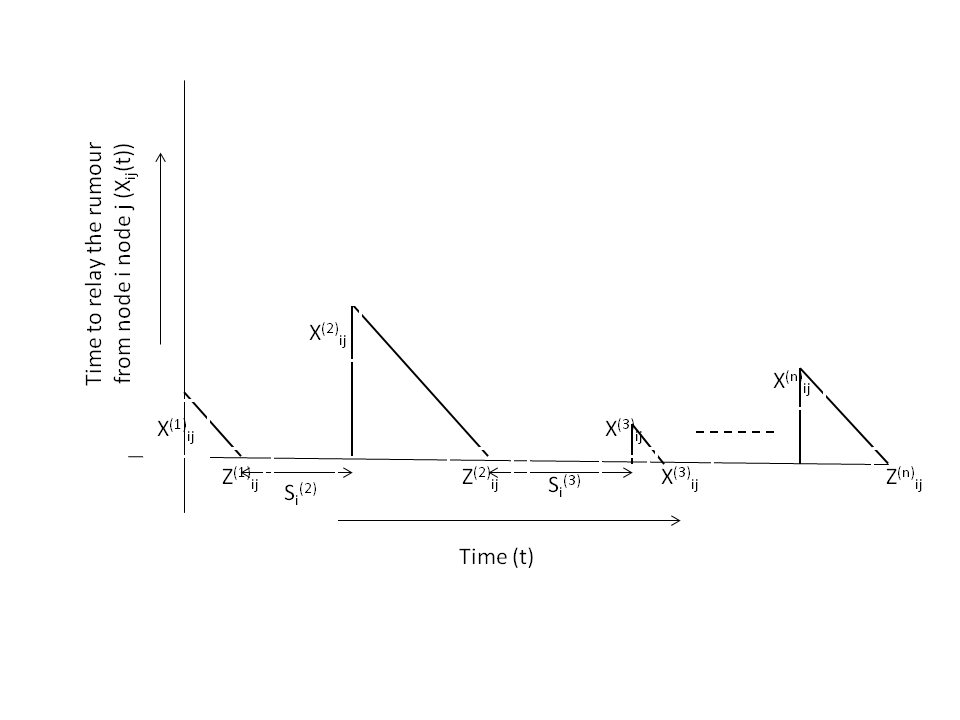}
	\caption{Timing diagram of the
process according to which sensor $i$ receives the
rumor  and disseminates it to sensor $j$. Sensor $i$ is idle
or inactive for times, $S_i^{(m)}$, $m\geq 2$ and in the
$m^{th}$ dissemination attempt, takes a time, $X_{ij}^{(m)}$
to actually reach node $j$ after the information is transmitted.    }
	\label{fig:timediagram1}
\end{figure}

The rumor dissemination process is depicted in
Figure \ref{fig:timediagram1} as a function of time.
The user corresponding to the $i^{th}$ sensor first
receives the information at time,
$Z_{ij}^{(1)}=X_{ij}^{(1)}$\footnote{Note that $Z_{ij}^{(1)}$ depends
only on $i$ and not on $j$. But we explicitly write $j$ to
be consistent with $Z_{ij}^{(m)}$, $m\geq 2$.}.
Then the $i^{th}$ sensor is idle for a period of time,
$S_2$ and then relays
the information at a time $X_{ij}^{(1)}+S_2$. The information
is received by the $j^{th}$ sensor at a time, $Z_{ij}^{(2)}=
X_{ij}^{(1)}+S_2+X_{ij}^{(2)}$. In general, the $i^{th}$ sensor
receives the information for the $m^{th}$ time at an epoch
$Z_{ij}^{(m)}$, stays idle for a time interval of length, $S_{m+1}$
and transmits the information at an epoch $Z_{ij}^{(m)}+S_{m+1}$,
which is received by the $j^{th}$ sensor for the $(m+1)^{th}$ time
at an epoch, $Z_{ij}^{(m)}+S_{m+1}+Z_{ij}^{(m+1)}$.This can be considered
as a \emph{renewal process with vacations} \cite{kleinrock}.\\
\begin{remark}
Note that the renewal process will not be
contiguous in time, as represented in Figure~\ref{fig:timediagram1}.
This is because, between the epoch when sensor $i$ receives the
information for the $(m-1)^{th}$ time and the $m^{th}$ time,
there is a time delay. However, that delay will not affect the
renewal theory-based analysis since we are interested in determining
the average time delay between the time epochs when the $i^{th}$ and
$j^{th}$ sensors receive the information. In other words, the \emph{blank}
period between the successive time instants the $i^{th}$ sensor
receives the information does not affect the renewal process.
\end{remark}

The time intervals, $X_{ij}^{(1)}$, $X_{ij}^{(2)}$, $\cdots$, $X_{ij}{(n)}$,
$\cdots$, are independent where $X_{ij}^{(1)}\sim A_{ij}$,
i.e., $\Pr\{X_{ij}^{(1)}\leq x\}=A_{ij}(x)$ and $X_{ij}^{(m)}\sim F_{ij}$,
i.e., $\Pr\{X_{ij}^{(m)}\leq x\}=F_{ij}(x)$, $m\geq 2$. Similarly,
$S_k$, $k\geq 2$ are independent and identically distributed
(iid) $S_k\sim V(x)$, $\forall$ $k$, i.e.,
$\Pr\S_k\leq x\}=V(x)$, $\forall$ $k$.
Let $a(x)=\frac{dA(x)}{dx}$, $f(x)=\frac{dF(x)}{dx}$ and
$v(x)=\frac{dV(x)}{dx}$, i.e., the probability density function (pdf)
of $X_{ij}^{(1)}$, $X_{ij}^{(m)}$, $m\geq 2$, and $S_k$, $k\geq 2$ are
$a_{ij}(x)$, $f_{ij}(x)$ and $v(x)$, respectively. \\

At any time epoch, $t$, let $Y_{ij}(t)$ be defined as the
\emph{remaining time} or \emph{residual transmission time}
of the information from sensor $i$ to the sensor $j$. Let
$\mathbf{Y}(t)=\left [ Y_{ij}(t)\right ]_{i,j\in\mathbf{S}}$;
$\mathbf{S}$ is the set of sensors. Then
in the analysis for the DN completion described in Section
\ref{subsec:DNcompletion}, specifically, we use $\lim_{t\rightarrow\infty}
E[\mathbf{Y}(t)]$,  instead of $E(\mathbf{X})$ in Eqns. (\ref{eqn:estalpha})
and (\ref{eqn:estbeta}).
The following theorems from renewal theory will be used to characterize
$E[\mathbf{Y}(t)]$.
\begin{theorem}\label{th:rrt}\cite{kleinrock}
Consider a renewal process
where the life time of the $l^{th}$ renewal is $X_l$. Let
$X_1\sim G(x)$, with pdf, $g(x)=\frac{dG(x)}{dx}$
and $X_2$, $X_3$, $\cdots\sim H(x)$, with pdf, $h(x)=\frac{dF(x)}{dx}$.
Let
\BEQA\label{eqn:defnmu}
\frac{1}{\mu}\define E(X)=\int_0^{\infty} [1-H(x)]dx =\int_0^{\infty} xh(x)dx.
\EEQA
Then the cumulative distribution function (CDF) of
$Y(t)$, $R(x)$ and the pdf of $Y(t)$, $r(x)=\frac{dR(x)}{dt}$,
are given by
\BEQA\label{eqn:CDFY}
\begin{array}{ccc}
R(x)&=&\lim_{t\rightarrow\infty}\frac{1}{t}\int_{u=0}^t \Pr\{Y(u)\leq x\}du\\
&=& \mu\int_{u=0}^x [1-H(u)]du,
\end{array}
\EEQA
\BEQA\label{eqn:pdfY}
r(x)=\mu [1-H(x)].
\EEQA
Moreover,
\BEQA\label{eqn:eofY}
E[(Y)]=\mu \int_{x=0}^\infty x^2 h(x)dx=\mu E(X_k^2), k\geq 2,
\EEQA
which, in turn, can be re-written as
\BEQA\label{eqn:eofYrewritten}
E[(Y)]=\frac{E(X_k)}{2} \left ( 1+C_X^2\right ),
\EEQA
where
\BEQA\label{eqn:skewness}
C_X^2=\frac{Var(X_k)}{\left [ E(X_k)\right ]^2}, k\geq 2.
\EEQA
\end{theorem}
\begin{theorem}\label{th:krt}\cite{kleinrock}
Let $K_y(t)\define
\Pr\{Y(t)\leq y\}$. Then
\BEQA\label{eqn:KRT}
\begin{array}{ccc}
\lim_{t\rightarrow\infty} K_y(t)&=&R(y),\\
\lim_{t\rightarrow\infty} [E(Y(t)]=\frac{E(X_k)}{2} \left ( 1+C_X^2\right ),
\end{array}
\EEQA
where $C_X^2$ is given by Eqn. (\ref{eqn:skewness}).
\end{theorem}
From Theorems \ref{th:rrt} and \ref{th:krt},
The average remaining time for information to reach from
sensors to each other, $E(\mathbf{Y})$, which we use in
Eqns. (\ref{eqn:estalpha}) and (\ref{eqn:estbeta})
is

\footnotesize
\BEQA\label{eqn:EofYUsedinDN}
E\left (Y_{ij}\right )=\frac{E\left (X_{ij}\right )+E \left (S_i \right )}{2}
\left ( 1+\frac{Var\left (X_{ij}\right )+Var\left (S_i\right )}
{\left [ E\left (X_{ij}\right )+E\left (S_i\right )\right ]^2}\right ).
\EEQA
 \normalsize

\noindent
In Eqn. (\ref{eqn:EofYUsedinDN}), $E\left (X_{ij}\right )+E\left (S_i\right )$
is the value $E(X_{ij})$ used in DN completion in Section
\ref{subsec:DNcompletion}, \emph{without applying the renewal theory-based analysis} discussed in this subsection. The expression for
$E\left (Y_{ij}\right )$ from Eqn. (\ref{eqn:EofYUsedinDN}) is
substituted in Eqns. (\ref{eqn:estalpha}) and (\ref{eqn:estbeta})
to obtain the modified DN completion using the renewal argument. The $\mbox{\boldmath{$\Delta$}}\mathbf{t}$ estimated in Sections \ref{subsec:compressedsensing}-\ref{subsec:renewal} are in used in the ML estimator described in Section \ref{sec:sourceid}, to identify the source of the rumor.

\section{Results and Discussion}\label{sec:results}

We conduct our experiments on the Sina Weibo dataset in \cite{weibo}.  Sina Weibo is the most popular
microblogging service in China \cite{weibowebsite}.
This dataset includes a followership network with $58,655,849$ nodes and $265,580,802$ edges, and a total of $370$ million tweets and retweets.
The retweeting paths (with their time-stamps) are  provided which is suitable in particular for studying  real information dissemination networks.  We selected $100$  tweets from this dataset  which constitute $100$ different real diffusion networks.
 Table~1 summarizes the details of the  dataset.
We used the Louvain method \cite{louvein}  to identify the   clusters, as    the gateway nodes of these clusters are used to construct the gateway graph $G^{\rm gate}$.
Since it is assumed  that rumors spread along the shortest paths into the social network, we selected nodes
 with high betweenness centrality as sensors.
\begin{table}
\caption{Details of dataset}
\label{Tab:data}
\centering
\begin{tabular}{|c|c|c|c|}
\hline
  &\multicolumn{1}{c|}{Max} & \multicolumn{1}{c|}{Ave} & \multicolumn{1}{c|}{Min} \\ \hline
\hline
Number of nodes & $43,545$&$41,978$&$40,445$\\\hline
Number of edges & $84,451$&$82,790$ &$80,923$\\\hline
Diameter &$13$ &$11$&$9$ \\\hline
Average shortest path length    & $8.31$&$5.97$&$4.71$\\\hline
Number of clusters  & $223$&$148$&$103$\\
\hline
\end{tabular}
\end{table}

\begin{figure}[htbp]
	\centering
		\includegraphics[width=0.5\textwidth]{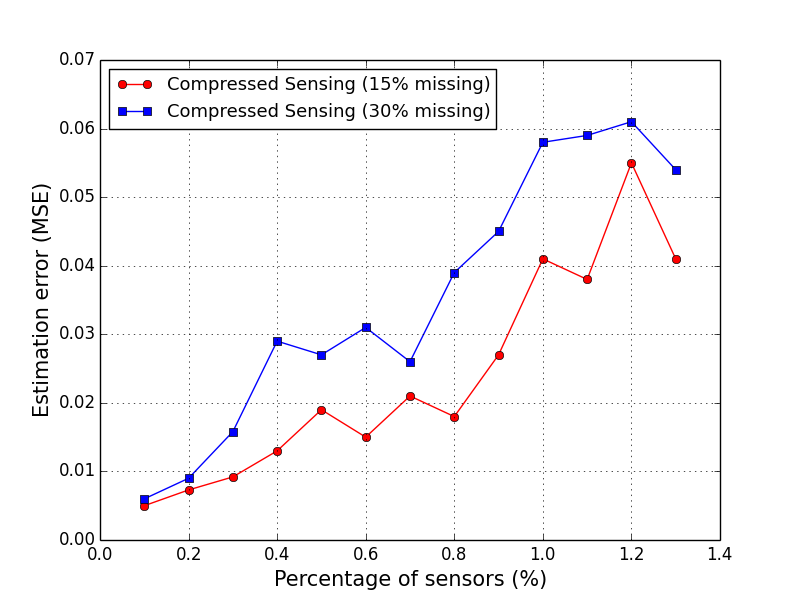}
	\caption{Estimation error for the observation vector $\mathbf{{\Delta t}}$
when $15\%$ and $30\%$  of entries are missing and deploying compressed sensing (described in Section
\ref{subsec:compressedsensing}).   }
	\label{fig:cs}
\end{figure}

    Figure~\ref{fig:cs}
		shows the accuracy of recovering the missing entries in $\mathbf{{\Delta t}}$ (employing compressed sensing) vs the percentage of
nodes used as sensors. The accuracy is  defined as
 the mean square error (MSE) between the original and the estimated missing entries. We randomly remove $15\%$ and $30\%$ of the entries  sporadically to simulate the  missing measurements.
As can
be seen from this graph, the estimation error is smaller when the missing rate
is smaller. It could be due to the fact that the number of remaining entries after $30\%$ missing  is less sufficient to precisely  estimate the missing entries.  However, the  error gap between the $15\%$  and  the $30\%$ missing rates  is very small
when the percentage of sensors is less than $0.3\%$.

\begin{figure}[htbp]
	\centering
		\includegraphics[width=0.5\textwidth]{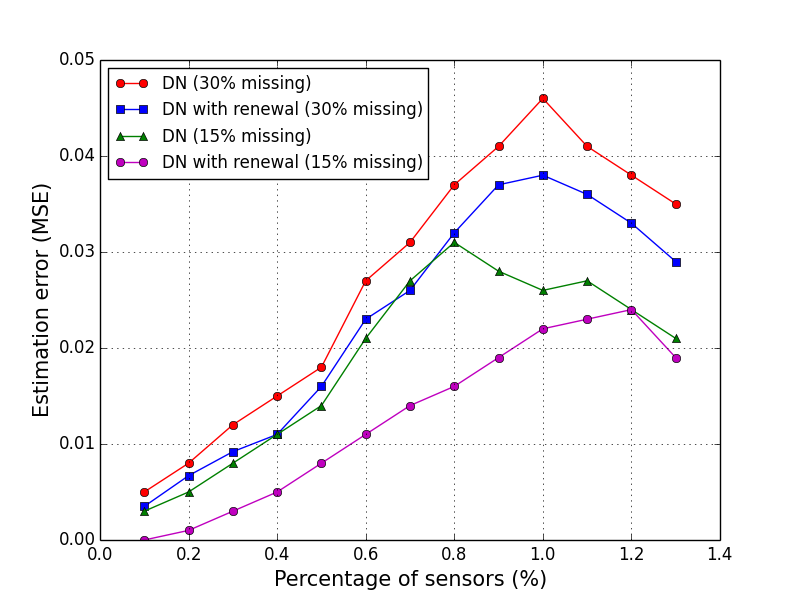}
	\caption{Estimation error for the observation vector $\mathbf{{\Delta t}}$ when the missing rate is $0\%$ (not missing measurements), $15\%$, and $30\%$ and deploying DN matrix completion (Section
\ref{subsec:DNcompletion}) and renewal based argument (Section
\ref{subsec:renewal}). The renewal theory based mechanism results in lower error because it utilizes the first two moments of the missing measurements while the DN completion method uses only the first moment.}
	\label{fig:dn}
\end{figure}

To evaluate the DN completion approach,  a sub-matrix of $D$ is removed. The removed sub-matrix is chosen such that the graph representation of the partial matrix forms a block clique.  Figure~\ref{fig:dn} shows the estimation error of the DN completion. The renewal based argument provides less estimation error because it utilizes the first two
moments of the dissemination intervals as opposed to the DN matrix completion method which uses only the first moment. It also shows that the accuracy improvement is larger when the missing rate is $15\%$.

\begin{figure}[htbp]
	\centering
		\includegraphics[width=0.5\textwidth]{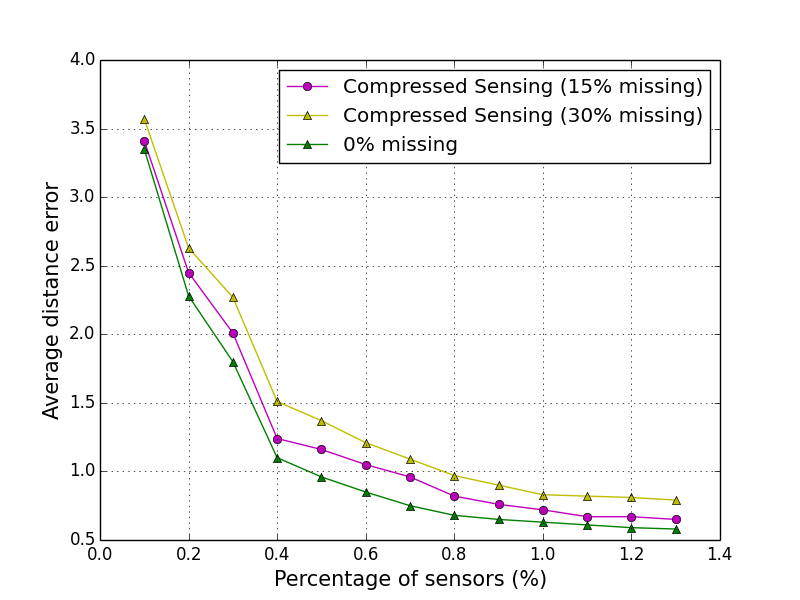}
	\caption{Average distance between the estimated
source and the actual source when $15\%$ and $30\%$
of entries in $\mathbf{{\Delta t}}$ are missing and  deploying compressed sensing (described in Section
\ref{subsec:compressedsensing}). }
	\label{fig:cs1}
\end{figure}

\begin{figure}[htbp]
	\centering
		\includegraphics[width=0.5\textwidth]{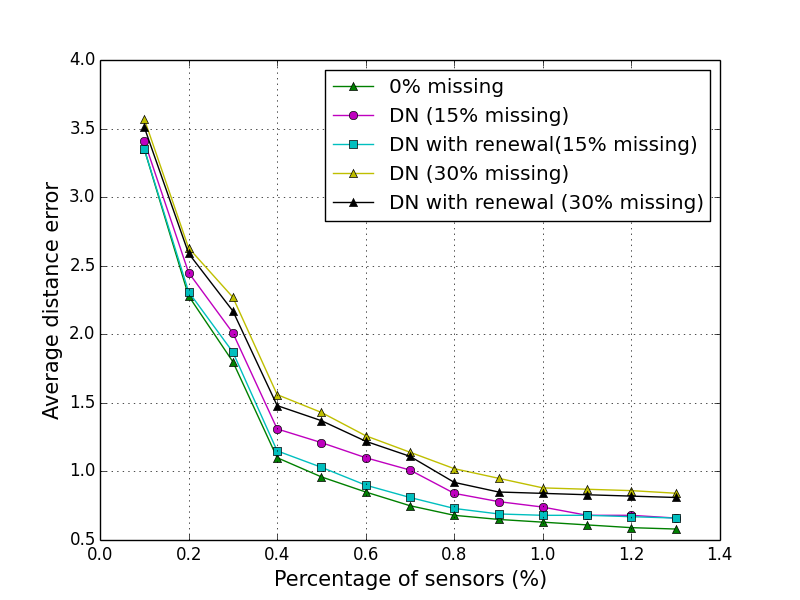}
	\caption{Average distance error  between the estimated
source and the actual source when $15\%$ and $30\%$
of entries in $\mathbf{{\Delta t}}$ are missing and deploying DN matrix completion (Section
\ref{subsec:DNcompletion}) and renewal-based argument (Section
\ref{subsec:renewal}). The renewal based argument provides less estimation error in source localization. }
	\label{fig:dn1}
\end{figure}

Next, we study the accuracy of the
 source estimation using   compressed sensing.
The accuracy is measured
in average distance between the estimated
 and the actual sources. Figure~\ref{fig:cs1} shows the source estimation error when the missing rate is $0\%$ (no missing measurements), $15\%$, and $30\%$.   It shows that compressed sensing results in almost as good a performance of the ML estimator as for
the network for which complete information is available. Figure~\ref{fig:dn1} shows the source estimation error when deploying DN matrix completion and renewal-based argument.  We again observe that the renewal based argument provides less estimation error in source localization. Results indicate that the estimation techniques result
in almost as good a performance of the ML estimator as for
the network for which complete information is available.

\section{Conclusions}
\label{sec:conc}
We addressed the problem of locating the source of a rumor in large-scale social networks with incomplete measurements. We presented the compressed sensing method to recover sporadically missing measurements and the doubly non-negative (DN) completion to recover  measurements missing in bursts. Furthermore, we presented a renewal theory-based model to boost the performance of the DN matrix completion method. We then used the recovered measurements to estimate the source of the rumor. We observed that  the compressed sensing and the DN matrix completion  provide less estimation error when the percentage of missing entries is less. It is also shown that the renewal theory-based model increases the accuracy improvement of the DN matrix completion method.  Mechanisms to jointly improve the ML estimator as well as the estimation of missing measurements, is under investigation.


\end{document}